\documentstyle{amsppt}
\magnification\magstep1
\hsize 16.5 truecm
\vsize 24 truecm
\parindent=1.5em                  
\parskip= 2pt plus 4pt
\voffset -1 truecm

\TagsOnRight

\define\a{\alpha}                      
\define\pd#1#2{\dfrac{\partial#1}{\partial#2}}
\define\pois#1#2{\{#1,#2\}}            

\define\R{{\Bbb R}}                    
\define\refno#1.{\par\smallskip\item{[#1]}\ignorespaces}
\define\w{\omega}                      
\define\W{\Omega}                      
\define\PC#1{\Theta_{#1}}              
\define\FL{{\cal F}L}                  
\def\g{{\gamma }}                   
\def\G{{\Gamma }}                   
\def\s{\sigma}
\define\z{{\theta}}
\define\e{\epsilon}                 
\define\la{\lambda}                     
\define\cv{{{\goth X}}}                 

\define\f(#1,#2){\frac {#1}{#2}}
\define\Lie#1{{\cal L}_{#1}}          

\define\bT#1{{\bold T}^{#1}}           
\define\bt{{\bold t}}                  
\define\T{\operatorname{T}}
\define\Ker{\operatorname{Ker}}        
\define\id{\operatorname{id}}          
\define\J#1{J^{#1}\pi}                 
\define\at#1{{}_{\big\vert_{\ssize #1}}}

\define\x{\times}
\define\cal{\Cal}

\define\SODE{{\eightpoint{SODE}}}
\define\dT#1{d_{\bT#1}}
\define\<#1>{\langle #1\rangle}

\def\fd{\hbox to 30pt{\rightarrowfill}}
\define\map{\longrightarrow}           


\define\sc{\smc}


\def\izab #1{\Big\downarrow \llap{$\vcenter{\hbox to 50pt{\hfil$\scriptstyle
#1$\ \ }}$}}
\def\izar #1{\Big\uparrow \llap{$\vcenter{\hbox to 50pt{\hfil$\scriptstyle
#1$\ \ }}$}}


\def\supobl #1{\llap{$\vcenter{\hbox to 15pt{$\scriptstyle #1$}}$}}

\overfullrule=0 pt
\def\Di50{1}
\def\BG55{2}
\def\MS74{3}
\def\CLR88{4}
\def\Ca90{5}
\def\FJ88{6}
\def\FlJ87{7}
\def\KMK90{8}
\def\KMKH92{9}
\def\FPS91{10}
\def\BNW92a{11}
\def\BNWC92b{12}
\def\CFM91{13}
\def\Ja94{14}
\def\Moa93{15}
\def\Mob93{16}
\def\Foocho{17}
\def\Fonueve{18}
\def\FouGre{19}
\def\Garri{20}
\def\MOT{21}
\def\BP{22}
\def\Ge02{23}
\def\Go{24}
\def\BN92{25}
\def\FZ{26}
\def\FRZZ{27}
\def\CF93{28}
\def\CLM91{29}
\def\PidTul{30}
\def\CFNM92{31}
\def\Crampin{32}
\def\CGuno{33}
\def\CGdos{34}
\def\Ab78{35}
\def\LR89{36}
\def\DS89{37}
\def\MC{38}
\def\BKMM{39}
\def\WSuno{40}
\def\WSdos{41}
\def\BK85{42}
\def\Fn98{43}
\def\Fa85{44}

\topmatter

\title
   On singular Lagrangians affine in velocities
\endtitle

\author
  Jos\'e F. Cari\~nena$^\dag$,
  Jos\'e Fern\'andez-N\'u\~nez$^\star$ and Manuel F. Ra\~nada$^\dag$
  \endauthor
\affil
  $^\dag$Departamento~de F\'\i sica Te\'orica, Universidad de Zaragoza,\\
  50009 Zaragoza {\smc(Spain)} \\
  $^\star$Departamento~de F\'\i sica, Universidad de Oviedo, \\
  33007 Oviedo {\smc(Spain)}
  \endaffil

\abstract
The properties of Lagrangians affine  in  velocities are analyzed in a
geometric way. These
systems are necessarily singular and exhibit, in general, gauge invariance.
The analysis of
constraint functions and  gauge symmetry leads us to a complete
classification of such
Lagrangians.
\endabstract

\leftheadtext\nofrills{J.F. Cari\~nena, J. Fern\'andez-N\'u\~nez and
M.F. Ra\~nada}
\rightheadtext\nofrills{On singular Lagrangians affine in velocities}
\endtopmatter

\document

\head
1. Introduction
\endhead
The Dirac's method of dealing with constrained systems \cite {\Di50-\MS74},
developed when looking for a way of quantizing such systems by means of
a \lq\lq canonical quantization\rq\rq-like procedure, has been
shown to be  a very efficient method and the geometric approach to
the theory of systems defined by singular Lagrangians has deserved a big
amount of attention during the past years.

In particular, Lagrangians that are affine in time-derivatives
have been analysed \cite{\CLR88} in the framework of presymplectic geometry
and the
geometric theory of time-independent singular Lagrangians \cite {\Ca90}.
They are important because their
Euler-Lagrange equations become systems of first-order
differential equations, appearing as constraints,
instead of  systems of second-order as it would happen with regular
  Lagrangians. So
they will play a relevant r\^ole
in many cases, not only in Physics, where many equations are first-order,
as in Dirac equation, but also
in other fields as in biology dynamics and Chemistry. On the
other side, these systems are singular and then they are
giving rise to gauge ambiguities and gauge symmetries.

The geometric study of a particular type of singular Lagrangian, those which
are affine in velocities,  was
carried  out in \cite {\CLR88} in the framework of  autonomous systems,
with the aim of studying the inverse problem of Lagrangian mechanics and
the theory of non-point symmetries from a new geometric perspective.
Almost simultaneously Faddeev and Jackiw developed a method for the
quantization of such singular systems which became soon very popular and
received
much attention from
most part of theoretical
physicists. This procedure of dealing with such systems  is usually referred
to as  Faddeev--Jackiw
method of
quantization \cite {\FJ88-\BNWC92b}.

One of the most important properties of singular Lagrangians is
  the existence of infinitesimal gauge symmetries which are related to
  the Second Noether theorem. This, which  is particularly important for
field theories, has only sense in the framework of time-dependent systems,
the time playing the r\^ole of base  coordinates in field theory.
A geometric
  approach to Second Noether Theorem was given in  \cite{\CFM91}, but
  the geometric theory of  the time-dependent description of such
affine in velocities Lagrangian systems
has never been developed, as far as we know,  even if  it is very important
as the only way of fully-understanding the meaning of  Noether's Second Theorem
establishing the relationship between singular Lagrangians and
gauge-transformations. Then
we feel  that a re-examination of the problem of singular Lagrangian that
are first-order in velocities will be very useful and will allow us a better
understanding of
the theory.

The two fundamental aspects of these systems described by
these first-order in velocities Lagrangians are the presence of
first order equations and constraints, which make possible this alternative
  Faddeev--Jackiw method of quantization \cite{\FJ88,\Ja94}.
This method is based on the reduction theory for  the presymplectic form
defined by the singular Lagrangian, but it admits  an alternative by means of
  the addition of the constraints with  some Lagrange multipliers in order
to obtain the symplectic
extension \cite{\BNW92a,\BNWC92b,\Moa93,{\Mob93}}. Once the symplectic
structure has been got and,
therefore, Poisson brackets have been defined, we can make use of  the
canonical quantization
procedure.
         In particular, the FJ-method uses  the reduction technique
and has been applied in many different fields, running form
condensed matter \cite{\Foocho-\FouGre}
and astrophysics \cite{\Garri}, until fluid dynamics \cite{\MOT} and, of
course, high-energy
  physics \cite{\BP}. Even the own
  Schr\"odinger equation can be derived using this method
\cite{\Ge02}.

It is also to be noted that the  FJ-method can be generalized to include
also fermionic degrees of freedom, i.e. non-commutative variables
\cite{\Go} and the corresponding canonical quantization can also be used in
super-symmetric theories \cite{\KMK90,\BN92,\FZ} with applications in
super-gravity (see
e.g. \cite{\FRZZ}).

The paper is then organized as follows. In Section 2 we summarize the
results of \cite {\CLR88}.
A Lagrangian approach to Hamilton  equations from a geometric perspective
will be  given in
Section 3. The framework for the geometric description of time-dependent
singular systems
as given in \cite {\CF93} will be indicated  in Section 4, where we will
also include a
recipe obtained from \cite{\CFM91} (see also  \cite{\CF93}) for the search
of gauge symmetries
for singular Lagrangian systems. The  geometric theory of time-dependent
  Lagrangians which are affine in the velocities will be developed in
Section 5 and the gauge
invariance of such systems
will be studied in Section 6. Finally, the theory is illustrated  with
several examples.

As a matter of notation, tangent and cotangent bundles will be denoted by
$\tau_M:\T M\to M$ and
$\pi_M:\T^*M\to M$, respectively. The set of vector fields  along a map
$f:M\to N$ (see e.g. \cite{\CLM91}), also called $f^*$-derivations in
\cite{\PidTul}, i.e. maps
$X:M\to\T N$ such that $\tau_N\circ X=f$, will be denoted  $\cv(f)$.
Examples of such kind of
vector fields are $\T f\circ Y$ and $Z\circ f$, where $Y\in\cv(M)$, $\T
f:\T M\to \T N$ is the
tangent  map corresponding to $f$, and
$Z\in\cv(N)$.
It has been shown in Pidello and Tulczyjew \cite{\PidTul} that a
vector field $X$ along
$f$ determines two $f^\star$-derivations of scalar forms on $M$: one of type
$i_\star$ and degree $-1$, denoted $i_X$, and other of type $d_\star$, denoted
$d_X$,
   defined in the following
way: given
$\a\in\bigwedge^p(N)$ and
$v_1,\ldots,v_{p-1}\in\T_xM$, we define  $i_X\a\in\bigwedge^{p-1}(M)$ by
$\<i_X\a(x),(v_1,\ldots,v_{p-1})>=\<\a(f(x)),(X(m),f_{*x}(v_1),\ldots,f_{*x}(v_{p-1}))>$,
and
$d_X\a\in\bigwedge^p(M)$ is given by
$d_X\a=di_X\a+i_Xd\a$  \cite{\CFNM92}.

\head
2. Geometric approach to time-independent  Lagrangians which are affine
in the velocities
\endhead

In the geometric description of a time-independent  Lagrangian system
the states are described by points of the tangent bundle $\tau_Q :\T Q\to Q$
of the configuration space $Q$, which is assumed to be a $n$-dimensional
differentiable  manifold.  We are interested in the geometric study of systems
described by a Lagrangian including only terms  up to first order
 in velocities,
namely, with a coordinate expression
$$
L(q,v)=m_j(q)\,v^j-V(q)\ ,\tag 2.1
$$
where summation on repeated indices is understood.

The property of a function being linear in the fibre coordinates
(velocities) of
the tangent bundle is intrinsic because it is preserved under point
transformations.  If $\phi\in C^{\infty}(\T Q)$ takes the form
$\phi=\phi_i(q)\,v^i$ in a particular  set of local coordinates,  then,
under the change of
coordinates $q^{\prime i}=q^{\prime i}(q)$, we have that in order to
$$
\phi_i(q)\, v^i\,=\,\phi_i^{\prime}(q^{\prime})\left (\pd{ q^{\prime i}}
{ q^j}\right )v^j
$$
the components $\phi_i(q)$ should transform as
$$
\phi_j(q)=\phi_i^{\prime}(q^{\prime})\left (\pd{ q^{\prime i}}
{ q^j}\right )\  , \qquad j=1,\dots, n\,,  \tag 2.2
$$
i.e.,  the functions $\phi_i(q)$'s should transform as the components of an
associated basic 1-form $\a\in\bigwedge^1(\T Q)$, $\a=\phi_i(q)\,dq^i$.
This suggests us  that there is thus a one-to-one linear correspondence
between basic 1-forms and
linear homogeneous functions that we establish next:
if $\mu\in \bigwedge^1(Q)$  is  a 1-form, then $\widehat \mu $ denotes the
function $\widehat \mu \in  C^\infty(\T Q)$  defined by $\widehat
\mu=i_{\bt}\mu$, where $\bt\in\cv(\tau_Q)$ is the identity map $\id:\T
Q\to\T Q$
viewed as a vector field along the tangent bundle projection $\tau_Q$; in
fibred coordinates, $\bt=v^i\,\partial/\partial q^i$. That is,
$\widehat\mu(q,v)=\langle\mu(q),\bt(q,v)\rangle$, the contraction making sense
because  $\bt(q,v)\in\T_qQ$. In coordinates,  when
$\mu= m_j(q)\, dq^j$,  $\widehat\mu(q,v)=m_i(q)\,v^i$.

 The
geometric theory for systems described by  a Lagrangian 
$L=\widehat\mu-\tau_Q^*V$,
with  $V\in  C^\infty (Q)$, that in coordinates of the bundle induced from
those
of  a chart in $Q$ becomes  $L=m_j(q)\,v^j-V(q)$, has been 
studied in \cite {\CLR88}: 
the energy $E_L$ and the presymplectic form $\,\w_L$ are  given by
$E_L=\tau_Q^*(V)$, $\theta_L=\tau_Q^* \mu$,  and, therefore,
$\w_L=-\tau_Q^*(d\mu)$, which in coordinates reads as follows:
$$\omega_L=\left(\pd{m_i}{q^k}-\pd{m_k}{q^i}\right)dq^i\wedge dq^k\ .
$$
   The Hessian matrix $W$ with elements
$W_{ij}=\partial^2L/\partial v^i\partial v^j$
vanishes identically and, therefore, all the $\tau_Q$-vertical vectors,
i.e. $\xi^i(q,v)\,\partial/\partial{v^i}$,  are
in the
kernel of $\w_L$.

The search for the other elements in the  kernel of $\w_L$ starts by
looking for a local
basis $\{Z_a=(z^i_a)\}$,  of the module of eigenvectors (if any)
corresponding to
  the null eigenvalue   of the matrix $A$
with elements given by
elements
$$
A_{ij}=\pd{m_j}{q^i}-\pd{m_i}{q^j}\ ,\qquad i,j=1,\dots, n\,. \tag 2.3
$$

Then, a basis for  the kernel of $\w_L$ is generated by
the vector fields
$X_a=z^i_a(q)\, \partial/\partial{q^i}$ and
$\partial/\partial{v^i}$, with   $a=1,\ldots,n-n_0$, where $n_0$ denotes the
rank
of the matrix $A$. The  primary constraint submanifold  $\Cal M$ is then
determined by the constraint functions  (see e.g. \cite{\Ca90})
$\phi_X=XE_L$, with $X\in\Ker\w_L$, that in the case we are considering are
$\phi_a=X_aE_L$, i.e.
$$
\phi_a=z^i_a(q)\,\pd V{q^ i}\, ,  \tag 2.4
$$
because the energy is $\tau_Q$-projectable, $E_L=\tau_Q^*V$, and then
$\partial{E_L}/\partial{v^i}\equiv 0$.

The general solution for the dynamical equation $i(X)\,\w_L=dE_L$ will be
given by
$$
X=\left[ \eta ^i+\lambda^a\,z^i_a\right]\pd{}{q^i}+f^i\, \pd{}{v^i}\ ,\tag 2.5
$$
with $\eta^i(q)$ being a solution of
$$
A_{ij}\,\eta^j=\pd V{q^ i}\, ,\tag 2.6
$$
and where $\lambda^a $ and $f^i$ are arbitrary functions on $\T Q$.

There is a special class of vector fields $D$ in $\T Q$ which are
called second order vector fields, hereafter shortened as  \SODE\
fields, which are characterized by  $S(D)=\Delta$, where $S$ denotes
the vertical endomorphism
\cite{\Crampin-\CGdos} and $\Delta$ is the Liouville vector field generating
dilations along the fibres.  
They can also be  characterized by
$\T\tau_{Q}\circ D=\bt$.

 The constraint functions for a Lagrangian given by (2.1)
are basic functions, i.e. holonomic
constraints,
defining a  submanifold $Q'$ of $Q$. Consequently, the secondary
constraints functions for the
existence of a solution restriction of  second order vector field  will be
$\widehat{d\phi_a}$ and
are given by linear functions in
the velocities.
A solution $X$ given by (2.5) can be the
restriction of a \SODE\ only in 
those points of $\T Q$ for which
$$
A_{ij}\, v^j=\pd{V}{q^i}\ .\tag 2.7
$$
In these points,  the general solution of the dynamical
equation is given by
$$
X=\left[ v ^i+\lambda^a\, z^i_a\right]\pd{}{q^i}+f^i\,\pd{}{v^i}\ ,\tag 2.8
$$
while  the \SODE\ condition corresponds to the choice $\lambda^a=0$.

The particularly simple  case in which $(Q,d\mu)$ is a symplectic manifold,
i.e. $\det A\not=0$, and therefore $Q$ is even dimensional, was also
studied in \cite {\CLR88},
where it was shown that then $\Ker \w_L$ is made of all $\tau_Q$-vertical
vectors and therefore there will be no dynamical constraint function. The
globally
defined solution of the  dynamical equation is
$$
X=\eta^i\,\pd{}{q^i}+f^i\,\pd{}{v^i}\ ,
$$
with the functions $\eta^i$ uniquely determined by
$$
\eta^i=(A^{-1})^{ij}\,\pd V{q^j}\ .\tag 2.9
$$

The Marsden--Weinstein theory of reduction for the presymplectic
system
defined by the Lagrangian (2.1),
$(\T Q,\omega_L=-\tau^*_Q(d\mu),E_L=\tau^*_Q(V))$, establishes that 
the reduced symplectic manifold is $(Q,-d\mu)$ with 
Hamiltonian function $V$.

The Hamiltonian formalism for the Lagrangian (2.1) was also studied in
\cite {\CLR88}: the primary
constraint  submanifold $P_1=\FL (\T Q)$ is determined  by
$\Phi_j(q,p)=p_j-m_j(q)=0$, i.e. is   the graph of the form $\mu$. Using the
identification of it with the base
$Q$ the pull-back of the canonical 1-form $\w_0$ in $\T^*Q$ is $-d\mu$.
The Poisson brackets of the constraint functions are  $\{\Phi_j,\Phi_k\}=
\{p_j-m_j(q),p_k-m_k(q)\}=A_{jk}$, and therefore, when $d\mu$
is symplectic, all the constraints are of the second class.
The Hamiltonian function  $H$ is defined
  on $P_1$   by the restriction
of the function $\widetilde
V=\pi_Q^*V$. The general
theory
leads again to the study of the  Hamiltonian dynamical system $(Q,-d\mu,V)$,
as in the Lagrangian case.

Let now $L$ be a singular Lagrangian for which all constraint
  functions 
$\phi_a=X_aE_L$, $a=1,\ldots , n-n_0=N$,  are holonomic; we
  consider an extended configuration space
$\R^N\times
Q$ and denote by ${\text{pr}}_2$ the natural projection
${\text{pr}}_2:\R^N \times Q\to Q$. We can then introduce  a
new Lagrangian $\bold L_1$ in
$\T(\R^N\times Q)$  by
$$
\bold L_1=\widetilde L+ \lambda^a\,\widetilde\phi_a\ ,\tag 2.10
$$
where the tilde stands for the
$\T{\text{pr}}_2$-pull-back and
$\lambda^a$
are the new additional coordinates (whose corresponding velocities will be
represented by $\zeta^a$). Taking into account that
$$\w_{\bold L_1}=\widetilde\w_L , \quad  E_{\bold
L_1}=\widetilde E_L-\lambda^a\,\widetilde\phi_a\ , \tag 2.11
$$
we see that $\Ker\w_{\bold L_1}$ is generated
by the set of  vector fields  projecting onto the vector fields  $X_a$  of
$\Ker\w_L$, plus $\partial/\partial \lambda^a$ and  $\partial/\partial
\zeta^a$.
The constraint functions determined by $\partial/\partial \lambda^a$ are
the (pull-back of the) original primary constraint functions $\phi_a$ and
\SODE\ condition leads us to consider the tangent bundle of the
new configuration space $\R^k\times Q'$.
The solutions of the dynamics will project under
${\text{pr}}_2$  onto the solutions of the original problem.

A similar  approach  can be done when we use $\widehat{d\phi_a}$ instead of
$\phi_a$ as constraint functions  and we  replace the original Lagrangian $L$
for $\bold L_2=\widetilde L+ \lambda^a\,\widetilde{\widehat{d\phi_a}}$. In
this case
$E_{\bold L_2}=\widetilde E_L$ and $\w_{\bold
L_2}=\widetilde\w_L+d\widetilde\phi_a\wedge
d\lambda^a$, from which we can see that
$$
X=\xi^i\pd{}{q^i}+\xi^a\pd{}{\lambda^a}+ \eta^i\pd{}{v^i}+
\eta^a\pd{}{\zeta^a}
$$
is in $\Ker\w_{\bold L_2}$ if and only if
$$
A\xi-\xi^a\nabla\phi_a-W\eta=0, \quad \xi\cdot\nabla\phi^a =0,\quad W\xi=0,
$$
from which we see that we will obtain as constraint functions (the
pull-back of)
those obtained directly from $L$, and the dynamics will correspond, up to the
gauge ambiguity in the coordinates  $\lambda^a$, to the dynamics  obtained in
$Q'$.

The relation $\lambda^a\,\widehat{d\phi_a}=\widehat{d(\lambda^a
\phi_a)}-\zeta^a\,\phi_a$ shows that the Lagrangian $\bold L_2$
may be replaced by $\bold L_3=\widetilde L-\zeta^a \,\widetilde\phi_a $,
which is
quite similar to $\bold L_1$ with the change of $\lambda^a$ for its velocity
$\zeta^a$.

In the more general situation for which $d\mu $ is singular, 
the primary constraint functions (2.4) will be holonomic and
therefore the previous remarks show us that they will determine a submanifold
$Q'\subset Q$ characterized by some constraint functions $\phi_a$ and the
corresponding secondary constraint functions  for a second order
evolution will be  $\widehat{d\phi_a}$. Every such a constraint can be
incorporated in a new Lagrangian ${\bold  L}$ defined in the tangent bundle of
a new configuration space ${\bold Q}$ of the form ${\bold  Q}=Q\times \Bbb
R^{(n-n_0)}$,    by
$$
{\bold  L}(q^i,\la,v^i ,\zeta) = L(q^i,v^i ) + \la^a \,\widehat{d \phi}_a
(q^i,v^i ),\tag 2.12
$$
where  $(q^i,\la^a,v^i ,\zeta^a )$  denote the coordinates on the tangent
bundle
$\T{\bold  Q}$.

The expressions for $\theta_{\bold L}$ and $\w_{\bold L}$ are
$$
\theta_{\bold L}= \widetilde\theta_L + \la^a \widetilde{d \phi_a}  \implies
\w_{\bold L}= \widetilde\w_L + \widetilde{d \phi_a}\wedge d\la^a , \tag 2.13
$$
and therefore  the rank of $\w_{\bold L}$ may be higher
than that of $\w_{L}$ and this is the starting point in the Faddeev--Jackiw
approach. The energy  $E_{\bold L}$ is the pull-back of $E_L$,
$E_{\bold L} = \widetilde E_L$.


\head
3. A Lagrangian approach to  Hamilton equations
\endhead
Let  $M$ be the configuration space of a mechanical
 system and consider  $Q=\T^*M$ 
endowed with its exact canonical symplectic structure $\w=-d\theta_0$,
where $\theta_0$ is the Liouville
1-form in $\T^*M$ (see e.g. \cite {\Ab78}). Then, given a function
$H\in C^\infty (\T^*M)$, let us define the linear Lagrangian $L\in  C^\infty
(\T(\T^*M))$ by
$$
L={\widehat \theta}_0-\tau_{\T^*M}^*H,\tag 3.1
$$
which in local coordinates is written
$$
L(q,p;v,u)=p_jv^j-H(q,p).\tag 3.2
$$

In this case $\theta_L\in  C^\infty (\T(\T^*M))$ is given by
$\theta _L=\tau_{\T^*M}^*\theta_0$ and $E_L=\tau_{\T^*M}^*H$.
The matrix $A$ given by (2.3) is now the symplectic matrix
$$J=\pmatrix 0&I\\-I&0\endpmatrix\ .
$$
The Kernel of $\w_L=-d\theta _L$ is then made up by the
$\tau_{\T^*M}$-vertical
vector fields
$$
Y_{f,g}=f^i(q,p;v,u)\, \pd {}{v^i}+g^i(q,p;v,u)\, \pd {}{u^i}\ ,\tag 3.3
$$
and therefore the presymplectic system defined by $L$, $(\T(\T^*M),\w_L,E_L)$,
admits a global dynamics which is not uniquely defined but given by
$$
\G_{f,g}=\pd H{p_i}\pd{}{q^i}-\pd H{q^i}\pd{}{p_i}+f^i(q,p;v,u)\, \pd
{}{v^i}+g^i(q,p;v,u)\, \pd {}{u^i}\ ,  \tag 3.4
$$
where $f$ and $g$ are arbitrary functions. The integral curves of each
one  of such
vector fields are determined by the system of differential equations
$$
\left\{\aligned \dot q^i&=\pd H{p_i}\\\dot p_i&=-\pd H{q^i}\\\dot
u^i&=f^i(q,p;v,u)\\\dot v^i&=g^i(q,p;v,u)\endaligned\right.
$$

However, only in those points for which
$$
\left\{\aligned v^i&=\pd H{p_i}\\u^i&=-\pd H{q^i}\endaligned\right. \tag 3.5
$$
the solution can be chosen to be the restriction of a \SODE\ vector field.
The preceding equations determine a submanifold $C$ on $\T(\T^*M)$ and
the condition on the restriction of  the vector  field $\G_{f,g}$ to be tangent
to $C$
determines the functions $f^i$ and $g^i$ by means of
$$
f^i=\G\left(\pd H{p_i}\right)\,,\qquad g^i=-\G\left(\pd H{q^i}\right)\,,
\tag 3.6
$$
obtaining in this way the vector field on $C$
$$\Gamma_C=\pd H{p_i}\pd{}{q^i}-\pd H{q^i}\pd{}{p_i}+
\G\left(\pd H{p_i}\right)\, \pd {}{v^i}-\G\left(\pd H{q^i}\right) \pd
{}{u^i}\ .  \tag 3.7
$$

The dimension of $C$ is only twice that of $M$ and then it can be
parametrized by $(q^i,p_i)$. The expression of $\Gamma_C$ in these
new coordinates is
$$\Gamma_C=\pd H{p_i}\pd{}{q^i}-\pd H{q^i}\pd{}{p_i}\ ,
$$
and therefore, here the Hamilton equations arise as determining the
integral curves of the uniquely defined vector field in the
submanifold $C$ in which such a \SODE\ solution of the dynamical equation
can be
found.  Then, a curve $\g: I\to M$ whose lift to $\T M$ lies in $C$ is the
solution
we were looking for.

\head
4. Geometric description of time-dependent singular system
\endhead

For the reader convenience we introduce in this Section the notation to be
used and a summary of
several properties and results of interest for
  the following sections.

The evolution space  of a time-dependent mechanical system whose
configuration space is the $n$-dimensional manifold $Q$ is $\R\x\T Q$ \cite{\LR89}, which  is
 the space of 1-jets of the trivial bundle $\pi\:\R\x Q\map\R$,
$j^1\pi=\R\x\T Q$; $\R$ is
endowed with a volume form $dt\in\bigwedge^1(\R)$ and represents the Newtonian
time. There is one  vector field in $\R$,  ${d/dt}\in\cv(\R)$,
   such that
$i({d/dt})dt=1$. The main geometric tools to be used in the geometric
description of
time-dependent mechanics are those of jet bundle geometry \cite{\DS89,\MC}. The $k$-jet
bundle of
$\pi$ is $J^k\pi=\R\x \T^kQ$, with $\T^kQ$ representing the space of
$k$-velocities. In particular, $J^2\pi=\R\x \T^2Q$ is the space of
accelerations
and $J^1\pi=\R\x \T Q$, the space of velocities.  By convention we will write
$J^0\equiv\R\x Q$. For each pair of indices
$k,l$ such that $k>l$,
there is a natural projection
$\pi_{k,l}\:\R\x \T^kQ\map\R\x \T^lQ$ and we will denote
$\pi_k=\pi\circ\pi_{k,0}\:\R\x
\T^kQ\map\R$, the projection of $J^k\pi$ onto $\R$.

If $\sigma\in{\hbox{Sec}}(\pi)$, then $j^k\s\in{\hbox{Sec}}(\pi_k)$
will denote  the $k$-jet
prolongation of $\s$. So, if $\s(t)=(t,\s^i(t))$, we have
$j^k\s(t)=(t,\s^i(t), {d\s^i/ dt},\dots, {d^k\s^i/ dt^k})$. We also recall
that the differential
1-forms $\theta\in\bigwedge^1(J^k\pi)$ such that $(j^k\s)^*\theta=0$, no
matter the
section $\s$, are called contact 1-forms of $J^k\pi$. They are the constraint
1-forms
for the so called Cartan distribution.

The theory of time-dependent Lagrangian systems makes an extensive use of the
notion of vector field along a map. In particular, there exist vector
fields along $\pi_{k+1,k}$, $\bT
k$,  representing the total derivative operators. They are
defined by means of
$\bT k\circ j^{k+1}\s=\T(j^k\s)\circ{d/ dt},\
\forall\s\in{\hbox{Sec}}(\pi).$ For every $F\in C^\infty(\R\x \T^kQ)$, the Lie
derivative of $F$ with respect to $\bT k$, i.e. the function
$d_{\bT k}F=i_{\bT  k}dF\in C^\infty(\R\x \T^{k+1}Q)$, represents the
  total time
derivative of $F$, usually written as ${dF/ dt}$ or simply $\dot F$. In fibred
coordinates $(t,q^i,v^i)$ for $\R\x \T Q$ and the corresponding ones,
$(t,q^i,v^i, a^i)$, for $\R\x
\T^2Q$, we have
$\bT 0={\partial/\partial t}+v^i\,{\partial/\partial q^i}\in\cv(\pi_{1,0})$ and
$\bT 1={\partial /\partial t}+v^i\,{\partial/\partial q^i}+
a^i\,{\partial/\partial v^i}\in\cv(\pi_{2,1}).$ Obviously $\bT 0$ and the
operator
$\bt\in\cv(\tau_Q)$ introduced in Section 2 are related, $\T\rho\circ\bT
0=\bt\circ\rho_2$, with
$\rho$ and
$\rho_2$ being the projections onto the second factor of $\R\x Q$ and
$\R\x\T Q$,
respectively.

A vector field $X\in\cv(\R\x Q)$ can be lifted to $\R\x \T Q$ giving rise
to a unique vector field
$X^1\in\cv(\R\x \T Q)$ which is $\pi_{1,0}$-projectable onto $X$ and
preserves the
set of contact 1-forms of $\R\x\T Q$. If the coordinate expression of $X$ is
$X=\tau\,{\partial/\partial t}+ X^i\,{\partial/\partial q^i}$, then
$$
X^1=\tau \,{\partial\over\partial t}+ X^i\,{\partial\over\partial q^i}+(\dot
X^i-v^i\,\dot\tau)\,{\partial\over\partial v^i}.
$$
Vector fields of type $X^1$ are
called infinitesimal contact transformations (hereafter {\sc ict});
it is worthy to
note that
$$
(fX)^1=(\pi_{1,0}^*f)X^1+\dot f\,X^V,
$$
where $X^V=S(X^1)$. Here  $S$ is the
vertical endomorphism on $\R\x \T Q$ \cite{\CFNM92}, which is a
(1,1)-tensor field whose
expression in
the fibred coordinates $(t,q^i,v^i)$ is
$S={\partial/\partial v^i}\otimes(dq^i-v^idt)$. Note that the local
1-forms given by
$\z^i=dq^i-v^i\,dt$ generate the set  of contact 1-forms of $\R\x\T Q$ and
the Cartan
distribution is but $\Ker S$.

A similar definition works for the prolongation of vector fields
$X=\tau(t,q,v)\,{\partial/\partial t}+X^i(t,q,v)\, {\partial/\partial
q^i}\in\cv(\pi_{1,0})$: the vector field along
$\pi_{2,1}$ given by
$$
X^1=\tau(t,q,v)\,{\partial\over\partial t}+X^i(t,q,v)\,{\partial\over\partial
q^i}+(\dot X^i-v^i\,\dot\tau){\partial\over\partial v^i}
$$
is  the {\it first prolongation of $X$}. Thus $\bT 1=(\bT 0)^1$.

More details about these notions and constructions can be found in
\cite{\CF93}.

The key object on which the geometric formulation of the dynamics
corresponding to a time-dependent Lagrangian $L\in C^\infty(\R\x \T Q)$
is based is the
Poincar\'e--Cartan 1-form, defined by
$$
\PC L=dL\circ S+L\, dt \in{\bigwedge}^1(\R\times \T Q).\tag 4.1
$$
In fibred coordinates $\PC L=(\partial L/\partial v^i)\,\z^i+L\,dt.$
Another important object related with $L$  is
  the Euler-Lagrange 1-form, which is defined by
$\delta L= i_{\bT 1}d\PC L\in{{\bigwedge}}^1 (\R\times \T^2Q)$, with local
coordinate expression $\delta L=L_i\,\z^i$, where
$$L_i=\frac d{dt}\left(\pd{L}{v^i}\right)-\pi_{2,1}^*\left(\pd
L{q^i}\right)$$ are the
variational derivatives of the
Lagrangian $L$.

The dynamical equation to be considered here is
$$
i(\G)d\PC L=0,\tag 4.2
$$
and, according to Hamilton's principle, the condition for the section
$\s\in{\hbox{Sec}}(\pi)$ to be an extremal of the action is
$$(j^1\s)^*[i(Z)d\PC L]=0\ , \qquad \forall Z\in\cv(\R\x\T Q)\ .$$
  The problem is to find a
vector field $\G\in\cv(\R\x\T Q)$ which is solution of the dynamical
equation (4.2) and
whose
integral curves are the first prolongation $j^1\s$ of sections
$\s\in{\hbox{Sec}}(\pi)$. In other words, $\G$ must be a time-dependent {\sc
sode}, i.e. a vector field $\G$ such that
$\T\pi_{1,0}\circ\G=\bT 0$. Its integral curves are parametrized  by $t$
and the local
expression of $\G$ is $\G= {\partial /\partial t}+v^i\,{\partial/\partial q^i}+
\G^i\,{\partial/\partial v^i}$. Both {\sc sode}-type and {\sc ict}-type fields
belong to $\Ker S$, but in general a {\sc sode} is not an {\sc ict}.

The submanifold of $\R\x \T Q$ where the dynamical equation
(4.2) possesses such
kind of solutions is given by the following theorem \cite{\CF93}:

\medskip

\proclaim{Theorem}
Let  $M_L$ denote the coisotropic subbundle
$$
M_L=\{Z\in {\T}(\R\x
\T Q)\mid S(Z)\in{\Cal V}_{1,0} (\Ker d\Theta_L) \}\ ,$$
and ${\Cal V}_{1,0} (\cdot)$ means the $\pi_{1,0}$-vertical part, which is but
the kernel of the differential of the Legendre transformation ${\Cal F}L\:\R\x
\T Q\map\R\x \T^*Q$, that is,
${\Cal V}_{1,0} (\Ker d\Theta_L)= \Ker\T{\Cal F}L.$

Then, there exist solutions of the dynamical equation which are
restrictions of a {\sc sode} field in the points, and only in that points, of
the set defined by  $$ {\Cal M}=\{z\in \R\x\T Q\mid
d\Theta_L(D,Z)(z)=0,\ \forall
Z\in
M_L,\ D{\hbox{ any {\sc sode} }}\}.
$$
\endproclaim
\medskip

When $L$ is regular, $M_L=\{0\}$ and ${\Cal M}=\R\x \T Q$ and,
consequently, there is no
restriction on the motion. But there exist
(primary) constraints for a singular Lagrangian which are given by the
following conditions
$$
\Phi_Z=d\Theta_L(D,Z)=0,\quad Z\in M_L. \tag4.3
$$
The functions $\Phi_Z$ are the {\it primary constraint functions}. Obvious
conditions for the consistency of the dynamics $\G$ compatible with the
constraints (4.3) are given by
$$
\chi_Z=\G(\Phi_Z)\vert_{{\Cal M}}=0\tag 4.4
$$
which either
give rise to the secondary constraints or (partially)
fix  the dynamics $\G$.
When the process is iterated, we will hopefully arrive at the final constraint
submanifold ${\Cal M}_f$, on which there exist solutions of the dynamical
equation which are the restriction onto
${\Cal M}_f$ of {\sc sode} fields tangent to ${\Cal M}_f$.

Note that here the only  ingredient is the singular Lagragian 
$L$ which 
provides both the (nonholonomic) constraint functions and 
the dynamics. 
In this sense this is a problem that  does not coincide
 with the more frequently studied constrained situation 
 in which
the starting point is a given nonholonomic 
constraint distribution \cite{\BKMM}. This latter situation, 
 that is receiving much attention during the last  years (see
e.g. \cite{\WSuno,\WSdos}), and is important by its relation with the theory of connections and by its 
control theoretical applications, is different from the 
present case of 
 first order singular  Lagrangians.

The dynamics obtained by applying the constraint algorithm sketched above
may be
not unique, a fact which is related with the gauge invariance of the
Lagrangian. The
appropriate geometric tool to deal with a gauge infinitesimal transformation
$$
\delta q^i=\e\,\sum_{\a=0}^R{d^{\a}g\over dt^{\a}}\,X^i_\alpha(t,q,v)\,,\qquad
\delta v^i={d\over dt}(\delta q^i)\,,\tag4.5
$$
where $g=g(t)$ is an arbitrary function of the time, is that of a vector
field along the bundle map $\pi_{1,0}$. In fact, let
$\{X_\a=X^i_\a(t,q,v)\,{\partial/\partial q^i}\mid \a=0,1, \dots,R\}$
be a family
of $R+1$ $\pi$-vertical vector fields along $\pi_{1,0}$ and $g(t)$ an
arbitrary function in $\R$. Then, if $X_g$ is the $\pi$-vertical vector
field along $\pi_{1,0}$
$$
X_g=\e\,\sum_{\a=0}^R{d^\a g\over dt^\a}X_\a\ ,\tag4.6
$$
its first prolongation $X_g^1$
is the infinitesimal generator of the gauge transformation (4.5).
Such $X_g$ is
said to be a gauge symmetry of the Lagrangian $L$ if there exists a function
$F_g\in C^\infty(\R\x\T Q)$ such that
$$
d_{X^1_g}L=X^1_gL=\frac{dF_g}{dt}\ .\tag4.7
$$
In such case
$\<\delta L,X_g>+{dG_g/ dt}=0,$ with $G_g=F_g-\<\PC L,X_g>$, and
conversely, if
there exists a function $G_g$ such that $\<\delta L,X_g>+{dG_g/ dt}=0$,
then $d_{X^1_g}L={dF_g/ dt}$ with $F_g=G_g+\<\PC L,X_g>$. (The contractions
$\<\delta L,X_g>$ and $\<\PC L,X_g>$ make sense because of the
$\pi_{2,0}$- and  $\pi_{1,0}$-semibasic character of $\delta L$ and $\PC L$,
respectively.)

All this mathematical apparatus can be used  to give a geometric
version of Noether's
Second Theorem as it can be seen in  \cite{\CFM91,\CF93}.
The theorem essentially establishes that a gauge-invariant Lagrangian is
necessarily
singular and it satisfies the so called Noether identities
$$
\sum_\a(-1)^\a\frac{d^\a\<\delta L,X_\a>}{dt^\a}=0\ ,
$$
i.e.
$$
\sum_{i,\a}(-1)^\a \frac {d^\a(L_iX_\a^i)}{dt^\a}=0\ ,
$$
with  $L_i$ being the variational
derivatives of the
Lagrangian $L$.  However, a singular Lagrangian needs not to be
gauge-invariant  and a method for
the determination of the gauge symmetry underlying a given singular Lagrangian
has been developed in
\cite{\CFM91} (see also \cite{\CF93}).
The method  is based on  the necessary conditions which are
derived from the second Noether's
theorem  and it also tells us whether, or not, a given Lagrangian is
gauge-invariant. Let  $X_g$ given by (4.6) be the wanted
gauge symmetry of $L$ and choose  a vector field $\widetilde X_{\a}\in\cv(\R\x
\T Q)$  such that
$\T\pi_{1,0}\circ \widetilde X_\a=X_\a$. Assume that
$$
G_g=\sum_{\a=0}^{R-1}G_\a\,\frac{d^{\a}g}{dt^{\a}}\ .
$$
 It follows from the symmetry condition (4.7) that the vector field
$X_R$ belongs to the
distribution $M_L-\Ker S$ and the functions $G_{\a}$ satisfy  the  relations
$$
G_R=0,\quad \dot G_{\a}+G_{\a-1}+\<\delta L,X_\a>=0 \quad
(\a=1,\dots,R),\quad \dot
G_0+\<\delta L,X_0>=0\,,
$$
from which it follows that all the functions $G_\a$ must be ${\Cal
F}L$-projectable and satisfy the
recursive relations
$$
G_{\a-1}=-d\Theta_L(D,\widetilde X_\a)-D(G_\a)\,,
$$
with $D$ being any {\sc sode}; in particular,
$G_{R-1}=-d\Theta_L(D,\widetilde X_R)$ is a
primary constraint function (see (4.3)). The  algorithm for the
determination of the gauge symmetries proceeds by determining in an iterative
and  orderly way the
functions $G_\a$ and the vector fields $X_\a$ along the following steps:

\item{\bf 1.} Choose
$G_R=0$ and select $\widetilde X_R\in(M_L-\Ker S)\cap\Ker\T\pi_{1}$ in such a
way that  $G_{R-1}=-d\Theta_L(D,\widetilde X_R)$, with $D$ being any {\sc
sode}, is a ${\Cal F}L$-projectable primary constraint function.

\item{\bf 2.} Then, let us  determine a $\pi_{1}$-vertical vector field $\widetilde
X_{R-1}$ in such a way that the 1-form
$\lambda_{R-1}=i(\widetilde X_{R-1})d\Theta_L-dG_{R-1}$ be
$\pi_{1,0}$-semibasic; $G_{R-2}$ is defined by  $G_{R-2}=\lambda_{R-1}(D)$.

\item{\bf 3.} When the successive functions $G_\a$ are ${\Cal
F}L$-projectable the
process may be iterated and  when we find
$G_{R-N}=G_R=0$ the algorithm enters into a cycle and the solutions appear
cyclically repeated. We can  take $R=N-1$ to be the higher order  for the
derivatives of $g(t)$ and the algorithm ends up.

In the case when in some step there is not any solution, we stop the
process and
return to make (if possible) a new choice for the solution in a previous step.
If there is no solution in any case, we have to conclude that the Lagrangian
is not gauge invariant.

The next sections are devoted to show the application of these constructions
to the case of a Lagrangian which is linear, or more accurately affine,
in the velocities $v^i$.

\head
5. Geometric theory of time-dependent Lagrangians which are affine in
velocities
\endhead

{}From the geometric viewpoint, a time-dependent Lagrangian $L$ which is
affine in the velocities arises from
a 1-form  $\lambda\in \bigwedge^1 (\Bbb R\times Q)$ in the following way:
$L=i_{\bT 0}\lambda  \in  C^\infty (\Bbb R\times \T Q).$
In fact, if $\lambda=m_i(t,q)\, dq^i-V(t,q)\, dt$, then the Lagrangian,
to be denoted by
$\widehat\lambda$, is
$$
\widehat\lambda=m_i(t,q)\, v^i- V(t,q)\ .\tag 5.1
$$
Obviously the
time-independent case we have dealt with in Sect. 2 is simply
a special case of this:
given  $\mu\in\bigwedge^1(Q)$ and $V\in C^\infty(Q)$ then we consider  the
1-form
$\la=\rho^*\mu-\rho^* V\,dt\in\bigwedge^1(\Bbb R\times  Q)$ which yields the
($\rho_2$-pull-back of the) time-independent Lagrangian (2.1).

Coming back to the general case, the basic geometric features for this
Lagrangian are:

\item{\bf 1.} The Poincar\'e--Cartan 1-form is given by
$\PC{\widehat\lambda}=\pi_{1,0}^*\lambda$.

\item{\bf 2.} ${\Cal F}\widehat\la$-projectability means
$\pi_{1,0}$-projectability, because
$${\Cal F}\widehat\la=\mu\circ
\la\circ\pi_{1,0}\ ,$$
where
$\mu=\pi_{\Bbb R}\x{\text {id}}_{\T^*Q}$ is the natural projection of
$\T^*(\Bbb
R\x Q)$ onto $\Bbb R\x\T^*Q$.

\item{\bf 3.} $\Cal V_{1,0}(\Ker d\PC{\widehat\la})=\Ker
\T\pi_{1,0}$, so that $M_{\widehat\la}= \T(\Bbb R\times \T Q).$

Taking into account all these facts, we find that the primary
constraint functions
are given by
$$
\Phi_Z= d\PC{\widehat\lambda}(D,Z)\,,\quad
  Z\in \T (\Bbb R\times \T Q),\quad D {\hbox{ any\  {\smc sode}}}\,,\tag 5.2
$$
i.e. $\Phi_Z=i(Z)(i_{\bT 0}d\lambda)$. The primary constraint manifold is
described in the following terms:
$$
\Cal M=\{z\in\Bbb R\times \T Q\mid \Phi_Z(z)=0\}\equiv \{z\in\Bbb
R\times \T
Q\mid i_{\bT 0}d\la(z)=0\}\ .\tag 5.3
$$
  However, if we recall that  there exists a local basis of
$\cv(\R\x\T Q)$ is made up from a {\smc sode} and vector fields  $Y^1$ and
$Y^V=S(Y^1)$, with  $Y\in\Ker \T\pi$ \cite{\CF93}, we see that the only
effective
constraint functions are those given by
$$
\Phi_Y= d\PC{\widehat\lambda}(Y^1,D),\ Y\in \Ker \T\pi\ .  \tag 5.4
$$
Then the functions $\Phi_Y$ are in a one to one correspondence with
the elements of $\Ker
\T\pi$. They  also  verify the property
$$
\Phi_{Y_1+fY_2}=\Phi_{Y_1}+(\pi_{1,0}^*f)\Phi_{Y_2}, \ f\in C^\infty(\R\x
Q)\ .\tag 5.5
$$
We will say that $\Phi_{Y}$ and $\Phi_{\overline Y}$ are `linearly dependent'
if $\Phi_{\overline Y}=\Phi_{fY}$ for some $f\in C^\infty(\R\x Q)$
everywhere non-null. This
property trivially takes place when ${\overline Y}=fY$ but in the case when
${\overline Y}$ and $Y$ are not dependent it is a property related with
the existence of a gauge
symmetry of the Lagrangian as we will see later.

In local coordinates $(t,q^i)$ for  $\R\x Q$ and  $(t,q^i,v^i)$ for
$\R\x \T Q$, respectively, the vector field $Y$ is written
$Y=Y^i(t,q)\,{\partial/ \partial q^i}$
and  $d\la=\frac12\,A_{ij}\,dq^i\wedge dq^j-\w_i\,dq^i\wedge dt$, where
$$
A_{ij}={\partial m_j\over\partial q^i}-{\partial m_i\over\partial
q^j}  \hbox{\rm \quad and\quad }
\w_i={\partial V\over\partial q^i}+{\partial m_i\over\partial t}.     \tag5.6
$$
Consequently, the primary constraint functions are
  $$
\Phi_Y=(A_{ij}\,v^j-\w_i)\,Y^i,\tag 5.7
$$
i.e. they are affine in the velocities. A basis for such
constraint functions is made up from the following functions
$$
\Phi_i=\Phi_{\partial/\partial q^i}=A_{ij}\,v^j-\w_i=0\,,
\qquad i=1,\ldots, n\,.\tag5.8
$$
Note that these equations are the Euler-Lagrange equations obtained from the
Lagrangian (5.1) and all of them appear in this formalism as constraint
equations.

  We can deduce from (5.4) that a primary constraint function  $\Phi_Y$ is
$\pi_{1,0}$-projectable (i.e., holonomic) if and only if  the  1-form
$i(Y)\,d\la$ is
$\pi$-semibasic. In an equivalent way, $\Phi_Y$ is $\pi_{1,0}$-projectable
if and only if
$$Y\in\W_{\la}=[\Ker\T\pi]^{\perp d\la}\cap\Ker\T\pi\ .$$
(The superscript ${\perp
d\la}$ means $d\la$-orthogonal complement).  In  local coordinates  as above,
$\W_{\la}$ is spanned by vector fields $Y=Y^i(t,q)\,{\partial/ \partial q^i}$
such that $A_{ij}\,Y^i=0$ and then the  corresponding constraint function
$\Phi_Y=\w_i\,Y^i=0$ is projectable.

Obviously, the maximum number  of linearly independent  constraint
functions equals the dimension $n$ of $Q$ and the maximum number of the
holonomic ones is $p=n-{\text{ rank }}A$,  where $A$ is the skew-symmetric
matrix of elements  $A_{ij}$ (5.6).

As far as the dynamics is concerned, any {\smc sode} $\G$ is a
solution on $\Cal M$  of the dynamical equation  because of (5.2). The
consistency  conditions (4.4),
$\G \Phi_Y=0$, will generate additional (secondary) constraints and/or fix
(maybe partially) the dynamics. More accurately,  the consistency
conditions for
non-projectable primary constraint functions  will determine some components of
the {\smc sode}, while the projectable ones will give rise to secondary
constraints that are affine in the velocities, which fix the dynamics. The
uniqueness is got when all the constraints so obtained  are independent and in
the maximum number. The two fundamental cases are.

I) The first fundamental case arises when $\W_{\la}=0$. Then the matrix $A$ is
regular and therefore there are no primary holonomic constraints. The $n$
independent constraints (5.8) are non-holonomic and lead to $n$ equations
determining in a unique way the dynamics on $\Cal M$, i.e.
$$
v^j=(A^{-1})^{jk}\w_k\ .\tag 5.9
$$
Of course, this is only possible when the dimension of $Q$ is
even, $n=2m$.

Let us analyse this ``regular'' case in geometrical terms. The condition
$\W_{\la}=0$ means that the distribution $\Ker d\la$ is 1-dimensional, so
$d\la$
is a contact form on $\Bbb R\x Q$ generating an exact contact structure on
$\Bbb
R\x Q$. In fact,  every nonzero vector field $Z\in\Ker d\la$ satisfies
  the condition $i(Z)dt\ne
0$, as it follows from the trivial fact that $Z$ also belongs to
$[\Ker\T\pi]^{\perp d\la}$, and if the condition $i(Z)dt= 0$ is fulfilled $Z$
should also be
$\pi$-vertical and, consequently, $Z=0$. Moreover for every pair $Z,Z'\in\Ker
d\la$ the vector field $Z''=[i(Z)dt]\,Z'-[i(Z')dt]\,Z$ lies in $\W_{\la}$,
$Z''\in \W_{\la}$, i.e.
$Z'=[i(Z')dt/ i(Z)dt]Z$.

On the other hand, no new constraints arise from the consistency conditions
(4.4)
and  the $n$ independent non-holonomic constraints will fix the dynamics on
$\Cal M$ in a unique way. Let $\eta\in\Ker d\la$ be  a vector field
such that
$i(\eta)dt=1$. It
generates locally $\Ker d\la$ and in coordinates turns out to be
$$
\eta={\partial\over \partial t}+(A^{-1})^{ij}\,\w_j\,{\partial\over\partial
q^i}\ .\tag 5.10
$$
The description (5.3) of
$\Cal M$ means that $\bT 0(z)\in \Ker d\la\,(\pi_{1,0}(z))$, $z\in
\Cal M$, and,
consequently, we can assert that the {\smc ict}
$\eta^1$ is, at least on $\Cal M$, a  {\smc sode} field. It is such
that
${i(\eta^1)d\PC{\widehat\la}}\vert_{\Cal M}=0$, thus
the dynamical equation (4.2) is satisfied by $\eta^1$ on $\Cal M$ and the {\smc
sode} $\G$ compatible with the constraints is the one generated by $\eta^1$.

In summary, when $\W_{\la}=0$, the Lagrangian system  on $\Bbb R\x\T Q$ reduces
to the (in general, time-dependent) Hamiltonian system  $(\Bbb R\x Q, d\la,H)$.
The manifold $\Bbb R\x Q$ is the extended phase-space and the Hamiltonian
function $H$ is essentially the energy $E_{\widehat\la}$, which is a
holonomic function:
$E_{\widehat\la}=\pi_{1,0}^*H$; in local coordinates $H=V$.

Both dynamical systems are equivalent in the sense that the integral curves
of the
dynamics in $\R\x\T Q$ are the first prolongation to $\R\x\T Q$ of the
integral curves
of the Hamiltonian dynamics in $\R\x Q$.

Using the 2-form $d\la$ we can define a Poisson bracket in $\Bbb R\x Q$
according to the following construction. Given a 1-form $\a\in\bigwedge^1(\Bbb
R\x Q)$ there exists a vector field $X_{\alpha}\in\cv(\Bbb R\x Q)$ such that
$i(X_{\alpha})d\la=\alpha$ iff $i(Z)\alpha=0$,
$\forall Z\in\Ker d\la$; obviously such a $X_\alpha$ is not unique, the
indeterminacy
being $\Ker d\la$ itself. Although $\alpha$ does not satisfies the condition
above one can take the ``$d\la$-semibasic'' part, given by
$$
\alpha^{\la}=\alpha-{i(Z)\alpha\over i(Z)\,dt}dt,\quad Z\in\Ker d\la\
,\tag 5.11
$$
which, in fact, depends only on $\Ker d\la$ and not on a particular $Z\in\Ker
d\la$, and
it trivially annihilates $\Ker d\la$, i.e. there exist vector fields
$X_\alpha$ such
that $i(X_{\alpha})d\la=\alpha^{\la}$.

In local coordinates, if $\a=a_0\,dt+a_i\,dq^i$, then
$$\a^{\la}=a_i\,\left(dq^i-(A^{-1})^{ij}\,\w_j\, dt\right)$$
and $$X_\a=X^0\,\eta+(A^{-1})^{jk}\,\a_j\,\pd{}{ q^k}\ ,$$ where
$X^0$ is an
arbitrary function and $\eta$ is given by (5.10); in particular, for $\a=df$,
$$d^{\la}f=(df)^{\la}=\pd f{q^i}\,(dq^i-(A^{-1})^{ij}\,\w_j\,
dt)$$ and
$$X_f=X_{df}=X^0\,\eta+ (A^{-1})^{jk}\,\pd f{q^j}\,\pd{}{q^k}\ .$$

The Poisson bracket $\{f,g\}$ of the functions $f$ and $g$ is then defined
by the rule
$$
\{f,g\}=d\la(X_g,X_f)=d^{\la}g(X_f)\ .\tag 5.12
$$
In local coordinates,
$$\{f,g\}=(A^{-1})^{jk}\,\pd f{q^j}\,\pd g{q^k}\,,$$ the fundamental
Poisson brackets
  being
$\{q^j,q^k\}=(A^{-1})^{jk}$ (see \cite{\BK85}). So the equation of motion
derived from the constraint equations (5.8) can be  written in the form
$\dot q^j=\{q^j,q^k\}\,\w_k$.

In the autonomous case $(Q,-d\mu)$ is a symplectic manifold and we obtain the
Hamiltonian system $(Q, -d\mu, V)$ analysed in detail in Section 2. In
particular, the equations of
motion read
$$\dot q^i=\{q^i,q^j\}\pd V{q^j}\ ,$$ (see (2.9)).

II) Second case: $\W_{\la}\not =0$. Then  $A$ is singular  and there
will be $r={\text{rank }}A$ independent non-projectable constraint functions,
and then  the number of effective projectable constraint functions is not
greater  than $p=n-r$. These functions generate secondary
constraint  functions that together with the non-projectable functions will
determine, at least partially, the dynamics. The uniqueness of solution
for the dynamics depends on whether  or not the primary
and  secondary constraints are independent. The analysis of gauge
invariance in the following section will clarify these points.

\head
  6.  Gauge  invariance of first order
Lagrangian systems
  \endhead

In this section we will show how to make use of
  the algorithm of  gauge  symmetry explained in Section 4.
It starts by taking $G_R=0$ and choosing a vector field  $\widetilde X_{R}
\in(M_{\widehat\la}-\Ker S)\cap\Ker\T\pi_1$ in  such a way that the function
  $G_{R-1}$ given by
  $G_{R-1}=d\Theta_{\widehat\la} (\widetilde
X_R,D)$ be a $\pi_{1,0}$-projectable  primary constraint function. Then, we
can  choose  $\widetilde X_R=Y^1$ (that is, $X_R=\T\pi_{1,0}\circ
Y^1=Y\circ\pi_{1,0}$,
where $Y\in\W_\la$), so that $G_{R-1}$ turns out to be
$G_{R-1}= \Phi_Y$ that is a $\pi_{1,0}$-projectable function.
Moreover, we can see
that  the  1-form  $\lambda_{R-1}=i(\widetilde X_{R-1})d\Theta_{\widehat\la}
-dG_{R-1}$ is
$\pi_{1,0}$-semibasic, no matter of the choice of
$\widetilde X_{R-1}$. We can write $\widetilde X_{R-1}$ as a sum
  $$
\widetilde X_{R-1}=Y^1_1+Y_2^V, \quad Y_1,Y_2\in \Ker\T\pi;\tag6.1
$$
and, consequently,
$$
G_{R-2}=d\Theta_{\widehat\la} (Y^1_1,D)-D(\Phi_Y)=\Phi_{Y_1}-\chi_Y\tag6.2
$$
is the difference between a primary constraint function and a secondary one.
The algorithm only works if it is possible to choose  $G_{R-2}$
as being a $\pi_{1,0}$-projectable function. The iteration of this
procedure will give rise to a sequence of
functions $G_{R-k}$ analogous to $G_{R-2}$, and therefore a gauge symmetry
will be obtained in
this way if we arrive at a secondary constraint function $\chi_Y$ that it
was a primary one,
making possible to have $G_{R-2}=0$.

Thus the starting point to have a gauge symmetry is to know the set of
holonomic (primary)
constraint functions. If two (or more) of them are linearly dependent it
will be possible to choose
$X_R$ in such a way that $G_{R-1}=0$. In the opposite case we need to know
whether or not the
primary constraint function gives rise to a secondary constraint function
which is a primary one.

More specifically, there are the following four possibilities and
correspondingly the four types of  affine in velocities
Lagrangians:

\bigskip

\noindent I) $\W_\la=0$, i.e.  $A$ is regular. In this case there are no
holonomic constraints and there is no gauge symmetry at all. The dynamics is
uniquely determined by the non-holonomic constraints (see Sect. 5).

\bigskip
\noindent II.1) $\W_\la\ne0$, i.e.  $A$ is singular, and there are two (non
trivially) linearly
dependent holonomic primary constraint  functions $\Phi_{\overline Y}$ and
$\Phi_{Y}$, i.e. $\Phi_{\overline Y}=\Phi_{fY}=(\pi_{1,0}^*f)
\Phi_Y$. In this case, $\overline Y-fY$ is non-null and choosing $\widetilde
X_{R}$ as  $\widetilde X_{R}=(\overline Y-fY)^1$, we will get
$G_{R-1}=0$ (i. e.,
the higher order for the derivatives of the arbitrary function $g(t)$ is $R=0$)
and  come to the gauge symmetry
$$
X_g=g(\overline Y-fY)\circ\pi_{1,0}.\tag6.3
$$

\bigskip

\noindent II.2) $\W_\la\ne0$  and there is a
holonomic primary constraint function $\Phi_Y$, $0\ne Y\in
\W_\la$, giving rise to a secondary constraint function
$\chi_Y=D(\Phi_Y)$ which is a  primary one, namely, there exists a
vector  field $\overline Y\in\Ker\T\pi$ such that
$\chi_Y=\Phi_{\overline Y}=d\Theta_{\widehat\la}
(\overline Y^1,D)$. In this case,
  choosing $Y_1=\overline Y$ we find that
$G_{R-2}=d\Theta_{\widehat\la} (\overline Y^1,D)-D(\Phi_Y)=0.$
Therefore the algorithm tells us that   $R=1$ and the  gauge symmetry
is
$$
X_g=g\,(\overline Y\circ\pi_{1,0})+\dot g\,(Y\circ\pi_{1,0})\ .\tag6.4
$$

\bigskip

\noindent II.3) $\W_\la\ne0$ and none of the secondary constraints is primary.
$G_{R-2}$ is not holonomic and the algorithm cannot go on, i.e. there is no
gauge symmetry. The dynamics is uniquely determined by the full set of
constraints (both primary and secondary).

\bigskip

In summary,  we will have  gauge symmetry when there exist
holonomic constraint functions generating  secondary constraint
functions  that generate a  free set with the non-projectable constraint
functions.

\head
7. Examples
\endhead

Finally, several  examples will be used  to illustrate the Lagrangian
analysis made in Sections 5 and 6. As a matter of notation we will use
subindices instead of
upperindices in the coordinates $q$ and the velocities $v$.

{\bf Example 1.} The well-known two-dimensional Lotka--Volterra system can be
derived from the  following Lagrangian which is affine
in the velocities \cite{\Fn98}:
$$
L=\frac{\ln y}{2\,x}\, v_x-\frac{\ln x}{2\,y}\,v_y-(a\ln y+b\ln x-x-y)\ ,
$$
where $a$ and $b$ are positive constants. Considered as a Lagrangian
function on
$\R\x\T\R^2_+$ it derives from the 1-form
$$
\la={\ln y\over 2x}dx-{\ln x\over 2y}dy-(a\ln y+b\ln
x-x-y)\,dt\in{\bigwedge}^1(\R\x\R^2_+)\ .
$$
Therefore,
$$
A=\pmatrix  0&-\dfrac 1{xy}\\ \dfrac1{xy} &0 \endpmatrix \quad{\text
{and}}\quad
\w=\pmatrix \dfrac bx-1\\\dfrac ay -1  \endpmatrix.
$$
The matrix $A$ is regular and then $\W_\la=0$. All of the primary
constraint functions
(5.8) are non-holonomic,
$$
\Phi_x=\Phi_{\partial/\partial x}=-{v_y\over xy}-{b\over x}+1,\qquad
\Phi_y=\Phi_{\partial/\partial y}={v_x\over xy}-{a\over y}+1,
$$
and, consequently, the reduced system (5.10) on $\R\x\R_+^2$ is
$$
\eta=\pd{}
t+x(a-y)\pd{} x-y(b-x)\pd{} y\ .
$$
  The system of differential equations
for its integral curves constitutes the 2-dimensional Lotka--Volterra system
$$
\dot x=x(a-y),\qquad  \dot y=-y(b-x)\ .
$$
  This is a Hamiltonian system with a symplectic
structure
$$
\sigma =\frac 1{x\,y}\,dy\wedge dx\ ,
$$
i.e. with defining Poisson bracket
$\{x,y\}=xy$, and Hamiltonian function
$$
H=a\ln y+b\ln x-x-y\ .
$$
  This Lagrangian is of Type I.

\bigskip

{\bf Example 2.} Another regular case is that provided by the Lagrangian
studied in \cite{\Fa85}
$$
L=\frac 12\,[(q_2+q_3)v_1-q_1v_2+(q_4-q_1)v_3-q_3v_4]-\frac
12\,(2q_2q_3+q_3^2+q_4^2)
$$
  coming from the 1-form $\lambda\in \in{\bigwedge}^1(\R\x\R^4)$ given by
$$
\la={1\over2}\big[(q_2+q_3)dq_1-q_1\,dq_2+(q_4-q_1)dq_3-q_3\,dq_4  \big]-
{1\over2}(2q_2q_3+q_3^2+q_4^2)\,dt\ ,
$$
from which we obtain
$$
A=\pmatrix 0&-1&-1&0\\ 1&0&0&0\\ 1&0&0&-1\\0&0&1&0\endpmatrix\quad {\text
{and}}
\quad\w=\pmatrix 0\\ q_3\\ q_2+q_3\\ q_4\endpmatrix.
$$
The matrix $A$ has rank 4 and then, as  $\W_\la=0$, there is not any
holonomic
constraint. The generating set of  primary constraints (5.8) is
$$
\Phi_1=-v_2-v_3=0\,,\quad
\Phi_2=v_1-q_3=0\,, \quad \Phi_3=v_1-v_4 -q_2-q_3=0\,,\quad
\Phi_4=v_3-q_4=0\ ,
$$
and the secondary ones, $\G(\Phi_i)=0$, with $i=1,\ldots,4$,
determine a unique dynamics which
is the
restriction of a {\smc sode} on the constraint manifold $\Cal M$, namely
$$
\G'=\G_{\vert\Cal M}={\partial\over\partial t}+q_3{\partial\over\partial
q_1}-q_4{\partial\over\partial q_2}+q_4{\partial\over\partial
q_3}-q_2{\partial\over\partial q_4}+v_3{\partial\over\partial
v_1}-v_4{\partial\over\partial v_2}+v_4{\partial\over\partial
v_3}-v_2{\partial\over\partial v_4}.
$$
The reduced system on $\R\x\R^4$ is
$$
\eta={\partial\over\partial t}+q_3{\partial\over\partial q_1}
-q_4{\partial\over\partial q_2}+q_4{\partial\over\partial q_3}
-q_2{\partial\over\partial q_4}\in\Ker d\la
$$

Note that the restriction of  $\G'$ onto $\Cal M$ coincides with that of
$\eta^1$.
\bigskip
{\bf Example 3.}  Let us now consider the Lagrangian of the type II.1
$$
L=q_1\,v_2+q_2\,v_3+q_2\,v_4-q_2\,(q_4-q_3)
$$
generated by the 1-form
$$
\la=q_1\,dq_2+q_2\,dq_3+q_2\,dq_4-q_2(q_4-q_3)\,dt\in{\bigwedge}^1(\R\x\R^4),
$$
from which we obtain
$$
A=\pmatrix 0&1&0&0\\  -1&0&1&1\\ 0&-1&0&0\\ 0&-1&0&0\endpmatrix\quad
{\text{and}}\quad
\w=\pmatrix 0\\ q_4-q_3\\ -q_2\\ q_2 \endpmatrix .
$$
The primary constraint functions (5.8) are
$$
\Phi_1=v_2\,,\quad
\Phi_2=-v_1+v_3+v_4-q_4+q_3\,,\quad
\Phi_3=-v_2+q_2\,,\quad \Phi_4=-v_2-q_2\ .
$$
The distribution $\W_\la$ is
spanned by the two
vector fields $\partial/\partial q_1+\partial/\partial q_3$ and
$\partial/\partial
q_1+\partial/\partial q_4$, and yields to two holonomic primary constraint
functions
$$
\Phi_1+\Phi_3=q_2\,,\qquad  \Phi_1+\Phi_4=-q_2
$$
that are linearly
dependent. The
Lagrangian is gauge invariant, and the symmetry vector (6.3) and the associated
function $F_g$ are ($R=0$)
$$
X_g=g(t)\Big(2{\partial\over\partial q_1}+{\partial\over\partial
q_3}+{\partial\over\partial q_4}\Big)\quad{\text{and}}\quad
F_g=2g(t)q_2\ .
$$
The dynamics $\G$ is determined by the  set of (primary and secondary)
constraints $\Phi_i,\ \chi_i$,  $i=1,\ldots,4$:
$$
\G={\pd{}{t}}+(v_3+v_4-q_4+q_3)\pd{}{q_1}+v_3\pd{}{q_3}+v_4\pd{}{q_4}+
(C_3+C_4-v_4+v_3)\pd{}{v_1}+C_3\pd{}{v_3}+C_4\pd{}{v_4},
$$
where $C_3$ and $C_4$ are arbitrary functions.
\bigskip
{\bf Example 4.} The Lagrangian defined by the 1-form
$$
\la=(q_2-q_3)\,dq_1-q_2\,dq_3-(q_2-q_1)\,q_3\,dt\in{\bigwedge}^1(\R\x\R^3)
$$
is of Type II.2. The matrices $A$ and $\w$ are given, respectively, by
$$
A=\pmatrix 0&-1&1\\ 1&0&-1\\ -1&1&0 \endpmatrix,\quad \w=\pmatrix
-q_3\\q_3\\q_2-q_1\endpmatrix.
$$
The set of primary  constraints (5.8) is generated by
$$
\Phi_1=-v_2+v_3+q_3\,,\quad
\Phi_2=v_1-v_3-q_3\,,\quad
\Phi_3=-v_1+v_2-q_2+q_1\ .
$$
  The rank of $A$ is 2 and therefore the distribution $\W_\la$ is
1-dimensional.
It is generated by the vector field $\partial/\partial
q_1+\partial/\partial q_2+\partial/\partial
q_3$ and there is one primary holonomic constraint function, namely,
$\Phi_1+\Phi_2+\Phi_3=q_1-q_2$,
whose corresponding secondary constraint function, $\chi=v_1-v_2$, is
likewise a primary
constraint:
$\chi=\Phi_{\partial/\partial q_1+\partial/\partial q_2}=\Phi_1+\Phi_2$.
Consequently,
this Lagrangian is gauge invariant: starting from $X_R=\partial/\partial
q_1+\partial/\partial q_2+\partial/\partial q_3$ and choosing
$X_{R-1}=\partial/\partial q_1+\partial/\partial q_2$ we obtain the gauge
symmetry (6.4)
$$
X_g=g(t)\left({\partial\over\partial q_1}+{\partial\over\partial q_2}\right)
+\dot g(t) \left({\partial\over\partial q_1}+{\partial\over\partial
q_2}+{\partial\over\partial q_3}\right)\ ,
$$
and the corresponding function $F_g=g(t)(q_1-q_3)-\dot g(t)q_3$.
The dynamics on $\Cal M$ is given by
$$
\G={\partial\over\partial t}+(v_3+q_3){\partial\over\partial
q_1}+(v_3+q_3){\partial\over\partial
q_2}+v_3 {\partial\over\partial q_3}+(C+v_3){\partial\over\partial
v_1}+(C+v_3){\partial\over\partial v_2}+C {\partial\over\partial v_3},
$$
with $C$ an arbitrary function.

\bigskip

{\bf Example 5.} The 1-form
$$\la=t\,q_2\,dq_1-q_1\,dq_2-[q_1-(t+1)q_2]\,q_3\,dt
\in\bigwedge\,^1(\R\x\R^3)
$$
gives the time-dependent Lagrangian of the type II.2
$$
\widehat\la=t\,q_2\,v_1-q_1\,v_2-[q_1-(t+1)q_2]\,q_3\ .
$$
The matrices $A$ and $\w$ are given, respectively, by
$$
A=\pmatrix 0&-(t+1)&0\\ t+1&0&0\\ 0&0&0\endpmatrix ,\quad\w=\pmatrix q_2+q_3\\
-(t+1)q_3\\ q_1-(t+1)q_2 \endpmatrix.
$$
Two  constraint functions of those determined by (5.8)
are non-holonomic
$$
\Phi_1=-(t+1)v_2-q_2-q_3\,,\qquad
\Phi_2=(t+1)v_1+(t+1)q_3\ ,
$$
  while  $\Phi_3$  is holonomic,
$$
\Phi_3=-q_1+(t+1)q_2\ .
$$
It gives rise to a secondary constraint,
$\chi_3=-v_1+(t+1)v_2+q_2$, which is a primary one,
$\chi_3=-\Phi_1-\Phi_2/(t+1)$.
The constraint function $\chi_3$ corresponds to the vector field
  $-\partial/\partial
q_1-(t+1)\partial/\partial q_2$ and, consequently, $\widehat\la$ is gauge
invariant. The
symmetry vector $X_g$ (6.4) and the function $F_g$ are ($R=1$)
$$
X_g=g(t)\Big(-{\partial\over\partial q_1}-{1\over t+1}{\partial\over\partial
q_2}\Big)+\dot g(t){\partial\over\partial q_3},\quad
F_g=g(t)\Big(q_2-{t\over t+1}q_1\Big).
$$

The local expression for the dynamical vector field is
$$
\G={\partial\over\partial t}-q_3{\partial\over\partial q_1}-{q_2+q_3\over t+1}
{\partial\over\partial q_2}+v_3{\partial\over\partial
q_3}-v_3{\partial\over\partial
v_1}-{2v_2+v_3\over t+1}{\partial\over\partial
v_2}+C{\partial\over\partial v_3},$$with $C$ being an arbitrary
function.

\bigskip

{\bf Example 6.} A Lagrangian of the type II.3 is
the one provided by a slight modification of the 1-form $\la$ in
Example
4:$$\la=(q_2-q_3)\,dq_1+q_2\,dq_3+(q_1-q_2)\,q_3\,dt\in{\bigwedge}^1(\R\x\R^3)\
.$$In this case,$$A=\pmatrix 0&-1&1\\ 1&0&1\\
-1&-1&0\endpmatrix,\qquad \w=\pmatrix -q_3\\ q_3\\
q_2-q_1\endpmatrix.$$
The primary constraint functions given by
(5.8) are$$\Phi_1=-v_2+v_3+q_3\,,\qquad\Phi_2=v_1+v_3-q_3\,,
\qquad \Phi_3=-v_1-v_2+q_1-q_2\ .$$
The distribution  $\W_\la$ is
spanned by the vector field\ 
$\partial/\partial{q_1}-\partial/\partial{q_2}-\partial/\partial{q_3}$.
There
is a holonomic constraint function,
$\Phi_1-\Phi_2-\Phi_3=2q_3-q_1+q_2$,
giving rise to a secondary one,
$\chi=2v_3-v_1+v_2$, which is not primary, that is, there is
no $\pi$-vertical vector field  $Y$ such that $\chi=\Phi_Y$. 
Consequently, there is no gauge symmetry at all, and the dynamics $\G$
on the
constraint manifold $\Cal M$ is
unique:$$\G=\pd{}{t}+q_3\Big(\pd{}{q_1}+\pd{}{q_2}\Big)+v_3\Big(\pd{}{v_1}+\pd{}{v_2}
\Big)\ .$$

\head
{Acknowledgements}
\endhead
The work of J.F.N.  has been partially supported by the
University of Oviedo, Vicerrectorado de Investigaci\'on, grant MB-02-514.
Support of Spanish DGI, BFM-2000-1066-C03-01  and FPA-2000-1252
projects, is also acknowledged.

\enddocument